\documentclass[twocolumn,showpacs,preprintnumbers,amsmath,amssymb,aps,prl,superscriptaddress]{revtex4}
\usepackage{graphicx}

\usepackage{bm}
\usepackage[usenames]{color}

\newcommand{\redsout}[1]{}
\newcommand{\red}[1]{#1}

\begin{document}

\title{Gain-induced trapping of microcavity exciton polariton condensates}

\author{Georgios Roumpos}
	\email{roumpos@stanford.edu}
	\affiliation{E. L. Ginzton Laboratory, Stanford University, Stanford, CA, 94305, USA}
\author{Wolfgang H. Nitsche}
	\affiliation{E. L. Ginzton Laboratory, Stanford University, Stanford, CA, 94305, USA}
\author{Sven H\"{o}fling}
    \affiliation{Technische Physik, University of W\"{u}rzburg Wilhelm-Conrad-R\"{o}ntgen-Research Center for Complex Material Systems, Germany}
\author{Alfred Forchel}
    \affiliation{Technische Physik, University of W\"{u}rzburg Wilhelm-Conrad-R\"{o}ntgen-Research Center for Complex Material Systems, Germany}
\author{Yoshihisa Yamamoto}
	\affiliation{E. L. Ginzton Laboratory, Stanford University, Stanford, CA, 94305, USA}
	\affiliation{National Institute of Informatics, Hitotsubashi, Chiyoda-ku, Tokyo 101-8430, Japan}

\date{\today}

\pacs{71.36.+c, 78.67.De, 03.75.Nt, 78.70.-g}

\begin{abstract}
We have performed real and momentum space spectroscopy of
exciton polariton condensates in a GaAs-based microcavity
under non-resonant excitation with an intensity stabilized laser.
An effective trapping mechanism is revealed, which is due to the
stimulated scattering gain inside the
finite excitation spot combined with the short lifetime.
We observe several quantized modes
while the lowest state
shows Heisenberg-limited real and momentum space distributions.
The experimental findings are qualitatively reproduced by an open dissipative
Gross-Pitaevskii equation model.
\end{abstract}

\maketitle

It is well known that an infinite two-dimensional (2D) system cannot establish true
long-range order above zero temperature because of phase fluctuations
\cite{Hohenberg1967,Mermin1966}.
But when a trapping mechanism is employed, so that the system size is finite,
a crossover temperature exists below which there are sizeable correlations
across the trap.
For example, a 2D system of ideal bosons in a trap is expected to show
Bose Einstein condensation (BEC) below a temperature that depends on the
trap size and shape \cite{Petrov2004}.

Microcavity exciton polaritons \cite{Weisbuch1992} is a 2D system that
behaves as bosonic at low densities.
Although the particle lifetime is short, limited by the photon lifetime inside the cavity,
so that true thermal equilibrium cannot be established,
a dynamical condensation phenomenon similar to BEC is observed
\cite{Deng2002,Kasprzak2006}.
Free electron-hole pairs are optically injected,
which then relax towards the lower polariton (LP) branch.
Above a threshold laser pumping power the scattering rate towards one or
a few final states is significantly enhanced due to the bosonic
final state stimulation, and condensation occurs.
It has been proposed that external trapping
such as photonic disorder \cite{Sanvitto2009}, stress trap \cite{Balili2007},
micropillar patterning \cite{Bajoni2008}, or metal film \cite{CWLai2007} is essential
to stabilize a spatial mode.

Here, we show that there is an implicit confinement mechanism
due to the finite lifetime and the spacially confined gain inside
the finite excitation spot size.
Namely, reservoir polaritons cannot travel far from the point where they were created
and final state stimulation only occurs inside the laser excitation spot area
where polariton density is sufficient.
Under suitable excitation conditions, we can observe the quantized
states of this effective trap in real as well as in momentum space.
Only the lowest state is perturbed by the disorder potential
and it is further confined by the latter.
We have observed Heisenberg-limited real and momentum space
distributions for this state, which confirms that phase fluctuations
are negligible.
We use a simplified one-dimensional model, employing a generalized
Gross-Pitaevskii equation with loss and gain terms, to qualitatively
reproduce our experimental results.

The sample is the same as in our recent experiments \cite{Roumpos2009,Roumpos2009b},
and it consists of an ${\rm AlAs}$ $\frac{\lambda}{2}$ cavity sandwiched between two
distributed Bragg reflector (DBR) mirrors.
The upper and lower mirrors are made of 16 and 20 pairs respectively of
${\rm AlAs}$ and ${\rm Ga_{0.8}Al_{0.2}As}$.
3 stacks of 4 ${\rm GaAs}$ quantum wells (QW's) are grown at
the central three antinodes of the cavity.
The measurements reported here are taken from a spot on the
sample with photon-exciton detuning $\delta=-2meV$, while the Rabi splitting
is $2\hbar\Omega_{Rabi}=14meV$.
The sample is kept at a temperature of $7-8K$ on the cold finger of a He flow
cryostat.
The system is pumped with a single-frequency Ti-Sapphire ring laser
focused on a flat spot of diameter $\sim 25\mu m$
(as measured for low pumping power)
from the direction normal to the sample \cite{Roumpos2009b}.
The laser is linearly polarized, and the orthogonal linear polarization is
detected.
Luminescence above threshold is almost unpolarized, namely the
degree of linear polarization is $<0.15$.
\red{This is attributed to the small ground state linear polarization splitting,
typically $\sim 50\mu eV$ in our sample \cite{Roumpos2009}.
We expect that the condensate randomly chooses one direction of linear
polarization in every realization, therefore luminescence appears
almost unpolarized after time-averaged detection.}
The laser wavelength is tuned to the first reflectivity minimum of the cavity
above the stop band and it is modulated into pulses with $0.5ms$ duration
and $100Hz$ repetition rate to prevent sample heating.
Luminescence is collected through an objective lens with numerical aperture
${\rm NA}=0.55$.
The spectroscopy setup is the same as in \cite{Roumpos2009}
and allows us to perform near field (NF - real space) and far field
(FF - momentum space) imaging and spectroscopy.
That is, we can measure energy-resolved luminescence as a function
of position or of in-plane momentum.
Our resolution is $1\mu m$ for NF spectroscopy
and $0.05\mu m^{-1}$ for FF spectroscopy.

\begin{figure}[t!]
\includegraphics[width=3.0in]{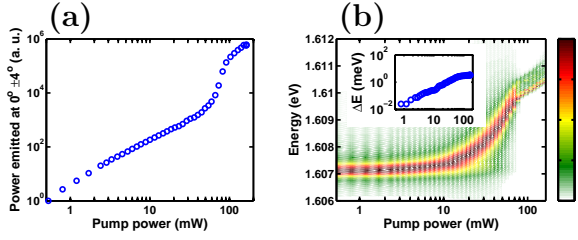}
\caption{(color online) Evidence for polariton condensation.
(a) Power emitted between $\pm 4^\circ$ as a function of pumping power.
A non-linear increase is evident at a threshold power of $70mW$.
(b) Normalized spectra as a function of power.
Above threshold, several narrow peaks appear.
Inset: Energy shift of the main spectral peak as a function of power.
The linear blue shift below threshold is followed by a logarithmic shift above threshold.
\label{fig:Condensation_Evidence}}
\end{figure}

\begin{figure}[t!]
\includegraphics[width=2.0in]{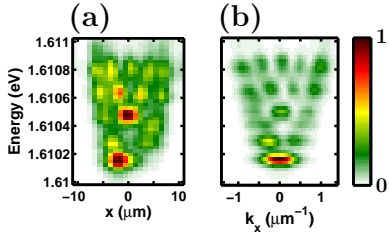}
\caption{(color online) (a) Real space and (b) momentum space spectra measured
above threshold (at $115mW$).
\label{fig:NF-FF_Spectra}}
\end{figure}

Above a threshold pumping power of $70mW$,
the phase transition from normal to condensation states is observed
(Fig \ref{fig:Condensation_Evidence}):
the power emitted at small angles increases nonlinearly,
spectral peaks with reduced linewidth appear
and the energy blue shift changes from a linear dependence on
the pumping power to a logarithmic dependence \cite{Bajoni2008}.
Fig. \ref{fig:NF-FF_Spectra} shows real space and momentum space
spectra measured above threshold.
Six discrete modes appear whose linewidth is resolution-limited.
Our spectral resolution has $HWHM\sim 25\mu eV$, corresponding to coherence time
$\hbar/HWHM = 26ps$.

\begin{figure}[t!]
\includegraphics[width=3.4in]{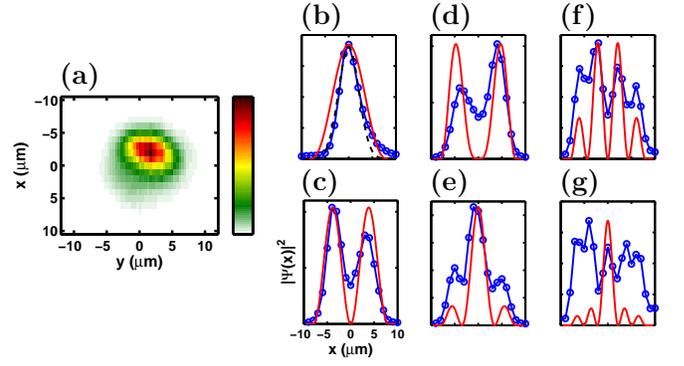}
\caption{(color online) (a) Experimental 2D real space image of the lowest trapped state
with quantum numbers $(m,n)=(0,1)$.
(b-g) 1D mode profiles. blue: experiment, red: trapped states
of an infinite circular quantum well of radius $8\mu m$.
In (b) we have integrated along the $y$-axis,
and also fitted the experimental profile with a gaussian distribution
(black dashed line).
(c-g) are simple cross sections along the $x$-axis at $y=0\mu m$.
The quantum numbers $(m,n)$ are
(b) $(0,1)$, (c) $(1,1)$, (d) $(2,1)$,
(e) $(0,2)$, (f) $(1,2)$, and (g) $(0,3)$.
\label{fig:NF_2DImages}}
\end{figure}

\begin{figure}[t!]
\includegraphics[width=3.0in]{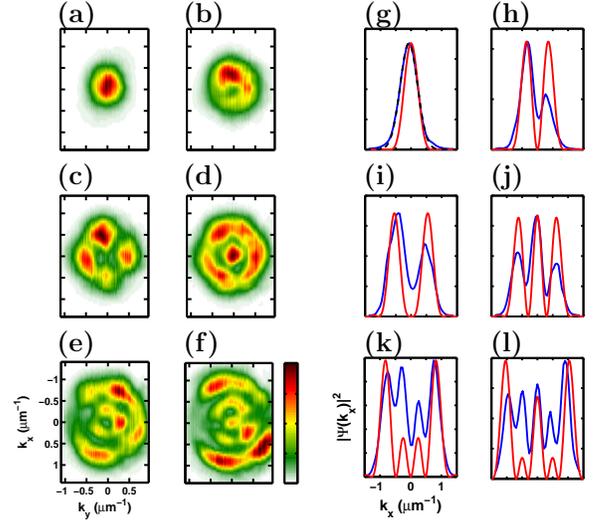}
\caption{(color online) (a-f) Experimental 2D momentum space images of trapped states.
The quantum numbers $(m,n)$ are
(a) $(0,1)$, (b) $(1,1)$, (c) $(2,1)$,
(d) $(0,2)$, (e) $(1,2)$, and (f) $(0,3)$.
(g-l) 1D mode profiles. blue: experiment, red: trapped states
of an infinite circular quantum well of radius $8\mu m$.
In (g) we have integrated along the $k_y$-axis,
and also fitted the experimental profile with a gaussian distribution
(black dashed line).
(h-l) are simple cross sections along $k_y=0\mu m^{-1}$.
The quantum numbers $(m,n)$ are
(g) $(0,1)$, (h) $(1,1)$, (i) $(2,1)$,
(j) $(0,2)$, (k) $(1,2)$, and (l) $(0,3)$.
\label{fig:FF_2DImages}}
\end{figure}

In order to move the real or momentum space image along one direction,
we move the last lens of our optical system (the one in front of the spectrometer)
along the same direction perpendicular to the optical axis.
The translation is also perpendicular to the direction of the spectrometer slit.
This way, we can record spectra along parallel lines in real or momentum space
resulting in a 2D grid of spectrally-resolved luminescence.
We can then integrate around the energy of each mode
and generate 2D images of the different modes in real (Fig. \ref{fig:NF_2DImages}(a))
as well as in momentum (Fig. \ref{fig:FF_2DImages}(a-f)) space.

Inside a circular trap with infinite barrier height, the solutions of the Schr\"{o}dinger
equation are expressed in terms of Bessel functions of the first kind
\begin{equation}
\psi_{mn}\left( r,\phi\right) \propto e^{im\phi}
{\rm J}_m\left( k_n r\right), \quad
k_n = \frac{j_{mn}}{R_0},
\end{equation}
where $R_0$ is the trap radius and $j_{mn}$ are the roots of the Bessel
function ${\rm J}_m(x)$ for $n=1,2,3\ldots$.
The FF wavefunctions follow from the Fourier transform of the above expression.
Based on this, we can assign quantum numbers $\left( m,n\right)$ to the
six observed modes as shown in Figs. \ref{fig:NF_2DImages}, \ref{fig:FF_2DImages}.

The real space images show the characteristics of the above wavefunctions.
The comparison is shown in Fig. \ref{fig:NF_2DImages}(b-g)
for a trap of radius $8\mu m$.
This radius is shorter than the spot radius measured below threshold
($\sim 12.5\mu m$), as the spot shrinks above threshold.
This is because the lifetime of the excitations we create drops from
$\sim 200ps$ below threshold to a few $ps$ above threshold,
\red{as measured for our sample and excitation conditions,}
so that they do not have enough time to diffuse.
\red{This is due to the bosonic final state stimulation effect,
namely the relaxation rate is proportional to the population of the final state,
so it is accelerated above the condensation threshold \cite{Deng2006}.}
For example, a polariton of mass $6\times 10^{-5}m_e$ and momentum
$k_x=0.5\mu m^{-1}$ can only move by $2\mu m$ in $2ps$.
The inability of polaritons to escape from the pumping spot
creates a finite-size gain region that traps condensates inside it.
This is a similar mechanism to gain guiding in semiconductor laser structures
\cite{Kapon1999}.

The same model works for the momentum space images, as shown in
Fig \ref{fig:FF_2DImages}(g-l).
We note that the harmonic potential model could not explain the energy
splitting between modes (2,1) and (0,2) (3rd and 4th modes)
which is clear in Fig. \ref{fig:NF-FF_Spectra}(b).
Also, the modes $(m,n)=$ $(3,1)$, $(4,1)$, $(2,2)$,
which should appear among the observed modes, are missing
since their large angular momentum makes polariton scattering into
these states inefficient.

The sample disorder potential follows a gaussian distribution with
standard deviation $<0.1meV$ \cite{Roumpos2009b}
which is considerably smaller than the kinetic energy of all modes
except the lowest one\red{, as shown on the measured spectra
of Fig. \ref{fig:NF-FF_Spectra}}.
Indeed, only the lowest mode is perturbed by the disorder potential
and it is further confined.
Fig. \ref{fig:NF_2DImages}(a) shows that the lowest mode has
migrated away from the center of the beam,
possibly towards a local minimum of the disorder potential.
In Fig. \ref{fig:NF_2DImages}(b) we plot the measured profile of this
mode along the $x-$axis, where we have integrated Fig
\ref{fig:NF_2DImages}(a) along the $y-$axis and shifted the result
to center it around $x=0\mu m$ (Instead, Figs.
\ref{fig:NF_2DImages}(c-g) are simple cross sections along $y=0\mu m$).
The observed distribution is narrower than the expected one
based on the infinite quantum well model.
We can fit it with a gaussian (dashed line) with standard deviation
$\sigma_x =2.16\mu m$.
The same effect is evident in the momentum space distribution of
the lowest mode (Fig \ref{fig:FF_2DImages}(g)) which is wider
than the model prediction.
Fitting with a gaussian gives a standard deviation of
$\sigma_{k_x}=0.283\mu m^{-1}$.
Taking the product $\sigma_x \times \sigma_{k_x}=0.61$,
which is very close to the Heisenberg limit of $0.5$.

The stabilized intensity of the excitation laser is critical for our observations.
In \cite{Roumpos2009b}, we used a mode-locked Ti-Sapphire laser operated
in the CW mode. As the laser cavity is designed for generating $psec$ pulses,
it supports a large number of modes with frequencies inside a
$ps^{-1}$-wide window,
so we expect power fluctuations at the $ps$ timescale.
Power fluctuations cause fluctuations in the polariton density
and the interaction energy is modulated accordingly.
This is the main decoherence mechanism \cite{Whittaker2009}.
Indeed, the coherence time of the polariton condensate measured
with a Michelson interferometer was $\tau_{coh}<5ps$ in that case.
In the present experiment, the laser cavity is designed for single-mode
operation and a double etalon is placed inside it to suppress all but
one mode.
The coherence time of the lowest observed polariton mode
is $\tau_{coh}>26ps$ as evidenced from the resolution-limitted linewidth in
Fig \ref{fig:NF-FF_Spectra}(b).
Measurement with the Michelson interferometer gives the result
$\tau_{coh}\sim 20ps$, but the real coherence time is expected to be
longer, as in this measurement we could not fully suppress the
higher modes.
This is consistent with the results reported in \cite{Love2008}
on a different sample.

We see that pumping with such a laser source creates the appropriate
conditions for a steady state condensate with long coherence time.
This is why only the eigenmodes of the gain-induced trap are visible
above threshold.
On the other hand, when the laser has power fluctuations on short
timescale, it introduces an energy uncertainty as explained above.
In this case, polaritons condense into a mixed state composed of
different modes closely spaced in energy.
In particular, phase fluctuations are prominent \cite{Roumpos2009b}.
The situation is more complicated in highly disordered samples
\cite{Krizhanovskii2009}, in which case the disorder potential plays a
big role in determining in which states polaritons condense.

The non-thermal population distribution among the different modes is
due to the non-equilibrium character of polariton condensation
and is determined by the pumping and loss rate of every state.
In particular, when we change the beam profile from a ${\rm \Pi}$-shape
to an M-shape, so that the pumping density is higher far from the center,
the lowest modes are suppressed.
The Heisenberg-limited NF and FF distributions of the lowest mode
confirm that this state is highly coherent, without substantial phase fluctuations
that would introduce energy uncertainty.
The interference patterns observed in the FF distributions of the
higher modes confirms the coherent character of these states, as well.

We employ a one-dimensional theoretical model
similar to the one used in \cite{Krizhanovskii2009}.
It cosists of two coupled equations, the open dissipative
Gross-Pitaevskii equation for the condensate order parameter $\psi(x,t)$
and the rate equation for the reservoir population $n_{R}(x,t)$
\begin{eqnarray}
i\hbar\frac{\partial\psi(x,t)}{\partial t}
&=& \big\{-\frac{\hbar^{2}\nabla^{2}}{2m^*}+V_{ext}(x)
-\frac{i\hbar}{2}\left[\gamma_{C}-Rn_{R}(x,t)\right] \nonumber \\
&+& g_{C}|\psi(x,t)|^{2} +g_{R}n_{R}(x,t)\big\}
\psi(x,t),\\
\frac{\partial n_{R}(x,t)}{\partial t}
&=& P_{las}(x,t) - \gamma_{R}n_{R}(x,t) \nonumber\\
&-& Rn_{R}(x,t)\left|\psi(x,t)\right| ^{2}.
\end{eqnarray}
$V_{ext}(x)$ is the external disorder potential and $\gamma_{C}$ is the
condensate decay rate.
For the stimulated scattering rate from the reservoir to the condensate
$R\times n_R(x,t)$, we use a linear dependence on the reservoir population.
$g_C$ and $g_R$ are the condensate-condensate and condensate-reservoir
interaction constants.
The reservoir density $n_R(x,t)$ is controlled by the laser pumping rate
$P_{las}(x,t)$ and the reservoir decay rate $\gamma_R$.
In the time-dependent simulation, $\psi(x,t)$ is a superposition of terms
oscillating at different frequencies, which correspond to different modes.

For our simulations, we set the interaction parameters $g_C$ and $g_R$
equal to zero, since when they are small enough they do not influence
appreciably the dynamics, but mostly only the energies of the
different modes in the steady state.
In this case, the stimulated scattering constant $R$ is irrelevant,
as a change in its magnitude can be absorbed by the reservoir population
$n_R(x,t)$ and the condensate order parameter $\psi(x,t)$ and results
in the same equations with a renormalized pumping rate $P_{las}$.
So, we can set $R$ constant and vary only $P_{las}$.
The threshold for condensation $P_{las}^{th}$ can be
calculated numerically, since below $P_{las}^{th}$ the solution
$\psi(x,t)=0$ is stable.
The other parameters we used are $m^*=6\times 10^{-5}m_e$,
$\gamma_C=(3ps)^{-1}$, $\gamma_R=(300ps)^{-1}$.

\begin{figure}[t!]
\includegraphics[width=3.0in]{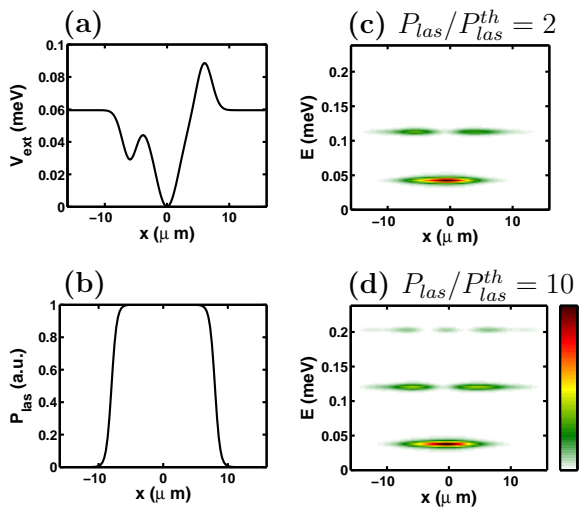}
\caption{(color online) Typical results of our theoretical model
(a) The disorter potential $V_{ext}(x)$ and
(b) the laser pumping rate profile $P_{las}(x)$ we considered.
(c-d) Real space spectra for different pumping powers,
2 and 10 times respectively above threshold.
Only the lowest mode is confined by the disorder potential.
\label{fig:Theory}}
\end{figure}

Our sample disorder potential is weak \cite{Roumpos2009b},
so the profile of Fig. \ref{fig:Theory}(a) is a resonable choice.
The results depend on the choice of this potential, but there is
no qualitative change as long as it has a local minimum
and it is not perfectly symmetric.
We also used a top-hat pumping profile with radius of $8\mu m$
(Fig. \ref{fig:Theory}(b)).
Fig. \ref{fig:Theory}(c-d) show the steady-state real space
spectra calculated under two different pumping powers
above threshold.
Excited states not confined by the disorder potential
spontaneously appear inside the area pumped by the laser,
in qualitative agreement with our experimental results.
A detailed comparison with the experiment requires a more
complete calculation in a two-dimensional grid,
and is beyond the scope of the paper.

We have presented evidence for an inherent effective trapping
mechanism of exciton polaritons, which is due to the finite size
of the excitation spot along with the short lifetime.
Under excitation with an intensity-stabilized laser,
several quantized modes of the effective trap spontaneously appear,
whose kinetic energies are higher than the disorder potential fluctuations.
We observed this quantized mode structure both in real
and in momentum space spectra and showed that the
real and momentum space distributions of the lowest
mode are Heisenberg limited.
A one-dimensional model based on the open dissipative
Gross-Pitaevskii equation with loss and gain qualitatively reproduces
our experimental findings.
This trapping mechanism should be present to all samples
studied so far.

\begin{acknowledgments}
This work was supported by
Navy/SPAWAR Grant N66001-09-1-2024, \red{MEXT,}
and by Special Coordination Funds for Promoting Science and Technology.
\red{W. H. N. acknowledges Gerhard Casper Stanford Graduate Fellowship.}
\end{acknowledgments}

\end{document}